\documentstyle[11pt,paspconf]{article}
\def\ltsim{ \,{}^<_\sim\, }

\begin{document}

\title{Issues in the Formation of Globular Cluster Systems}

\author{William E.~Harris}
\affil{Department of Physics \& Astronomy, McMaster University, Hamilton
    ON L8S 4M1, Canada}


\section{Cluster Ages in the Milky Way Halo}

Globular clusters 
make up less than 1\% of the visible halo stars.  These, in turn,
form only a few percent of all the mass present
in the potential well of a large galaxy. 
However, these
ancient star systems contain information about the history of halo
formation out of all proportion to their fraction of the total mass.
The literature on globular cluster systems in the Milky Way and
other galaxies is now far too large to summarize in one short review,
so I will concentrate only on a few recent issues that
have taken on particular importance.

Within our own Milky Way, globular cluster {\it ages} have once again
taken center stage in the debate over halo formation.
With continuing in stellar physics and evolution codes, 
the best estimates
of absolute ages for the oldest globular clusters have undergone modest
downward revisions.  A very considerable factor helping to drive them
in this direction has been the appearance of the {\it Hipparcos} parallaxes,
with much new information on field-star subdwarfs used to
calibrate globular cluster distances (e.g., Reid 1997; Gratton et al.
1997; Pont et al. 1997).
This new material, though important, has not yet settled the distance
calibration issue; early reports of resulting cluster ages as low as
10 Gyr are not being borne out by subsequent studies, and it may
not be too early to suggest an emerging new consensus of theory
and observation around  $\tau \simeq (13 \pm 1)$ Gyr for
the oldest, most metal-poor halo clusters.  

\begin{figure}
\plotfiddle{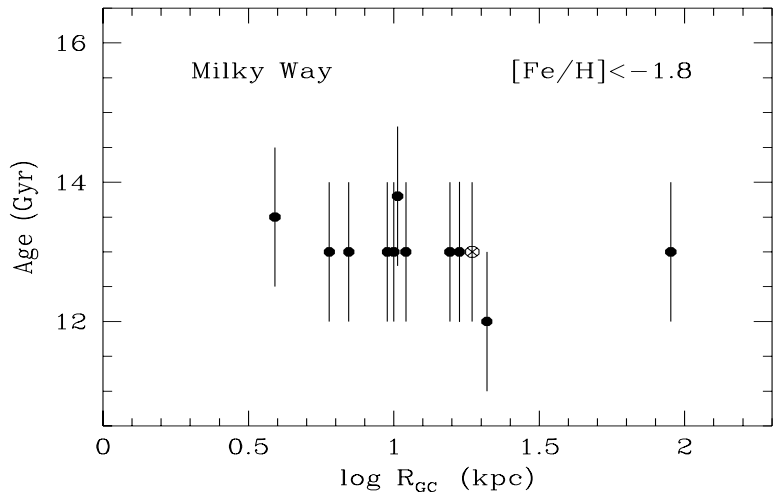}{2.50in}{0}{60}{50}{-185}{-110}
\caption{Ages of low-metallicity ([Fe/H] $\ltsim -1.8$) globular
clusters in the Milky Way; data from Harris et al. (1997a) and
Richer et al. (1996).  The zeropoint of the age scale is adopted
as 13 Gyr (see text).  The circled cross denotes M54, the
largest and most metal-poor globular cluster in the Sgr dwarf
(Layden \& Sarajedini 1997). Relative age uncertainties 
average $\pm1$ Gyr per cluster (cf. Harris et al. 1997a).} \label{fig-1}
\end{figure}

Independently of absolute age calibrations, we can estimate
the {\it duration} of the halo formation era by finding  
{\it relative} cluster ages from widely different parts of the halo
and over the full range of cluster metallicities.  With present-day
high precision main sequence photometry, these relative
ages can be gauged to within $\pm 0.5$ Gyr for clusters of similar
compositions.  Various recent discussions (Richer et al. 1996;
Chaboyer et al. 1996; Bruzual et al. 1997)
suggest that there is an overall age-metallicity relation, with
the more metal-rich clusters (47 Tuc and the others like it)
being $\sim 2$ Gyr younger on average than the most metal-poor
(and presumably oldest) ones like M92 and M15.  
But perhaps surprisingly, we are discovering 
that the ages of the {\it lowest}-metallicity clusters
are identical to within measurement precision everywhere in the halo
from Galactocentric distances of $\sim 5$ kpc out to $\sim 100$ kpc
(see Figure~\ref{fig-1}).  
The most recent and telling comparisons of this type
are by Harris et al. (1997a, for the remote-halo cluster
NGC 2419 versus the inner-halo clusters of similar metallicity).

The implications is that,
all over the protogalaxy which would eventually evolve into the
Milky Way, globular cluster and halo-star formation began at 
virtually the same time.  This volume is now almost 200 kpc 
in diameter.  How can we fit this in to the overwhelming 
evidence for a lumpy formation history and the buildup of the
halo from many pieces (see below)?  This upper end to
the cluster age distribution may simply be telling us that any of 
the pregalactic `pieces' (dwarf-sized $10^8 - 10^9 M_{\odot}$ gas clouds)
could have begun building stars at about the same time after they emerged from
the recombination era and began collecting into the protogalactic
potential wells.  These protodwarfs (or SGMC's; see
Harris \& Pudritz 1994; McLaughlin \& Pudritz 1996) 
are exactly the right environments for the buildup of 
protoglobular clusters, a process which should take
a few $\times 10^8$ years, regardless of location within the halo.

High precision new calibrations of the relative ages of clusters,
based firmly on deep main-sequence photometry, are continuing
to accumulate.  Within a few years the long-sought goal of a complete
{\it age profile} for the Galactic halo may be within our grasp.
With it may come the first unequivocal solution to the famous
`second-parameter' problem affecting the color-magnitude distributions 
of stars in the diagrams of the metal-poor clusters, which have long
been suspected to be due primarily to age differences (e.g. Lee et al. 1994).

\section{The Deconstructionist Halo}

\begin{figure}
\plotfiddle{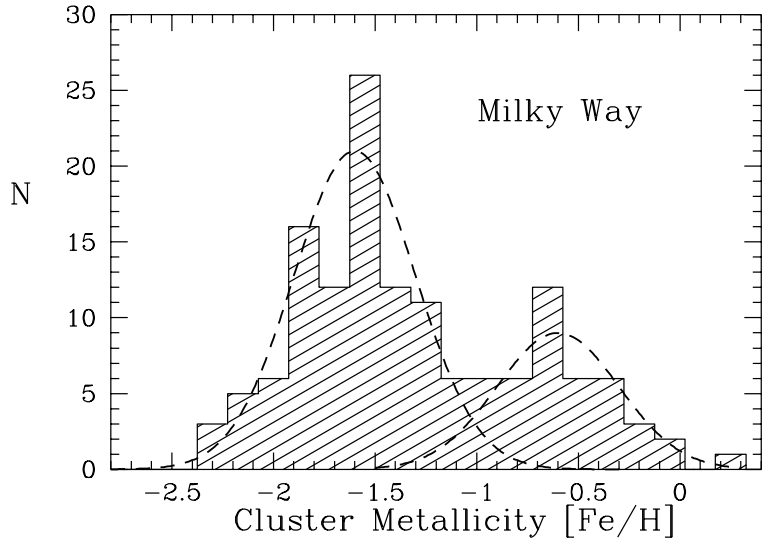}{3.00in}{0}{70}{70}{-195}{-135}
\caption{Metallicity distribution for all globular clusters in
the Milky Way, with recent data from Harris (1996).  Gaussian curves
are shown with $\sigma$[Fe/H] = 0.3 and peaks at [Fe/H] $ = -1.6,
-0.6$.} \label{fig-2}
\end{figure}

Kinman (1959) first demonstrated that the Milky Way globular cluster
system contained at least two kinematically distinct subgroups which
also differed in mean metallicity.
The landmark paper of Zinn (1985) strongly reinforced this conclusion 
with considerable new data, and related it
in modern terms to the probable formation histories of the different
metallicity groups.   The case for two major subpopulations builds 
from what is now seen to be a remarkably
common phenomenon of halo cluster systems in other galaxies:  a {\it bimodal
metallicity distribution} (Figure~\ref{fig-2}).  
In the Milky Way, the clusters more metal-poor
than [Fe/H] $\sim -1.0$ define what we traditionally think of as the ``normal halo'',
with a Gaussian-like metallicity distribution centered at [Fe/H] $ = -1.6$
and little or no systemic rotation, closely resembling the field halo stars.
By contrast, the metal-richer subpopulation ([Fe/H] $> -1.0$) carries a 
significant overall rotation, and (if the best current age estimates are correct)
may be younger on average by up to 2 Gyr than the metal-poor halo.
Zinn (1985) and Armandroff (1989) favored interpreting these clusters
as a disk system, possibly identified with the stellar thick disk,
and most workers have referred to them collectively as 
``disk clusters'' since.  However, this 
identification is very much open to debate.  Minniti (1995)
has pointed out that the kinematics of the metal-richer clusters, coupled with
their rather restricted space distribution in the innermost few kpc, may
be more consistent with a bulge-like population rather than the thick disk.
\begin{figure}
\plotfiddle{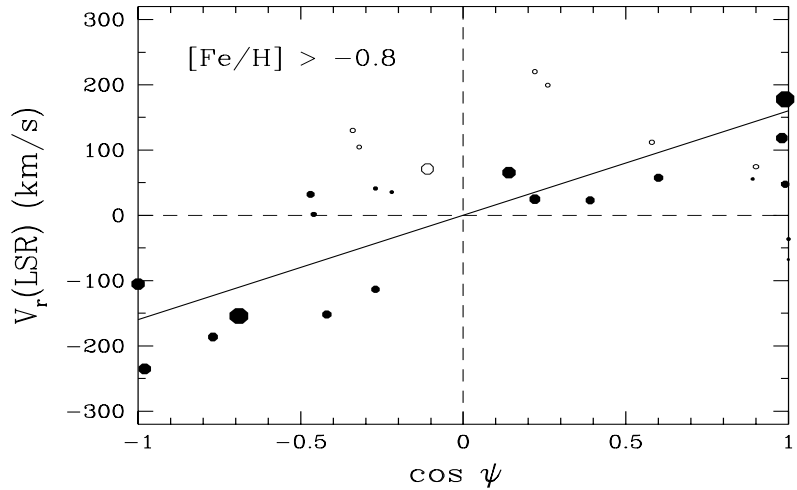}{2.60in}{0}{70}{60}{-200}{-140}
\caption{Radial velocities for metal-rich globular clusters in the
Milky Way, with data from Harris (1996).  Here $v_r$ is the velocity
relative to a stationary point at the Sun; in this graph, a constant
$v_{rot}$ appears as a straight line (see Zinn 1985 for definition 
of the angle $\psi$).  The solid line is $v_{rot} = 160$ km/s.
Higher-weight points are plotted with larger
symbols; the open circles are those clusters claimed
by Burkert \& Smith (1997) to be in a barlike configuration.}
\label{fig-3}
\end{figure}

Figure~\ref{fig-3} shows the kinematical distribution.  
The small number of objects, the internal scatter, and the effects of
distance errors for these highly reddened clusters
(which make the actual positions of many points in the graph
highly uncertain and thus of low weight) would allow any $v_{rot}$ 
within the range of $\sim100$ to 200 km/s, which could be either
bulge-like or disk-like.
To cloud the issue further, Burkert \& Smith (1997) claim
that some of these inner clusters (those with the 
most positive radial velocities
and smallest longitudes) may form a bar-like system embedded in
the bulge.  Unfortunately these same 
objects tend to be among the ones with the worst
known distances and reddenings. 

Whether or not the clusters truly represent any subpopulation
of the field {\it stars} is, once again, closely connected with ages.
Holtzman et al. (1993) estimate that the bulge stars 
have a mean age near 10 Gyr on a scale where the inner globular clusters are 
13 Gyr.  A very similar $\sim 3-4$ Gyr difference between field stars 
and clusters is found by Layden \& Sarajedini (1997) for the Sagittarius 
dwarf elliptical.  If the Milky Way halo was built from the amalgamation 
of many such pieces, the evidence suggests 
that the clusters had a considerable
head start, and were present for a long time while most of the
Galaxy's raw material was still gaseous.  Earlier arguments 
(see Harris 1991) based on space distributions, free-fall times, 
and metallicity enrichment timescales, had suggested
that globular clusters might be 
generically older by perhaps $\sim1$ Gyr than the
bulk of the halo in most large galaxies -- quite a bit less than
the direct age differences now emerging from the new color-magnitude studies.

Attempts have also been made to deconstruct the metal{\it-poor}
cluster population into distinct subgroups.  Morphological differences in
the color-magnitude diagrams certainly correlate with galactocentric distance
(the second-parameter problem; e.g. Lee et al. 1994), 
and may correlate with age and kinematics.  One suggested subgroup may even
have a retrograde net rotation, and it is at least possible that it may
represent a particularly large satellite accretion event
(Rodgers \& Paltoglou 1984; Zinn 1993; van den Bergh 1993).
Better and more direct age information should help to evaluate the range
of possibilities.

The importance of the cluster/field-star age offset is considerable.  
There appears to have been 
plenty of time after the initial burst of globular cluster formation for
the leftover and newly enriched gas -- which in fact 
was {\it most} of the Galaxy`s raw material -- 
to mix, dissipate, recollect into new clouds, and recollapse before beginning
new and more substantial bursts of star formation that gave rise to the bulge,
the thick disk, and eventually the thin disk.  By that time, much of its memory of the 
initial state from the cluster-forming epoch would have been thoroughly
erased as it evolved into different subpopulations.  Thus the long-standing question
whether or not the globular clusters can be taken as 
`representative' of some major stellar population in the bulge or halo seems
to me to be moot -- essentially unaswerable.  At present, 
they represent only themselves.
To be sure, some fraction of the field-star component must consist of stars
tidally stripped from existing clusters, or from small clusters that were entirely
dissolved; but current observations, simulations, and theoretical constraints
suggest that this fraction is not likely to exceed $\sim10$\% (e.g.,
Harris \& Pudritz 1994; Capriotti \& Hawley 1996; Murali \& Weinberg 1997).

\section{Giant E Galaxies:  Stormy Beginnings}

Observations for globular cluster systems in other galaxies
have been advancing by leaps and bounds in the
last few years.  Accurate metallicity distributions for huge samples of
clusters are starting to become available for key objects such as NGC 4472,
NGC 1399, M87, and the other gE's in Virgo and Fornax (e.g., Whitmore 
et al.~1995; Geisler et al.~1996; Forbes et al.~1997a,b).  
With these come more comprehensive luminosity functions
(GCLFs) and space distributions, and eventually (with the 8-meter class of
telescopes; cf.~Cohen \& Ryzhov 1997) spectra and
dynamical analyses.  From these studies, two common themes have emerged
so strongly that they must take precedence in any respectable scenario of
cluster formation:  

\noindent (1) The near-universality of the {\it mass distribution
function} (GCLF):  in galaxies of all types and sizes, the clusters follow
$dN/dM \sim M^{-1.8 \pm 0.3}$, except for the lowest $\sim10$\% of the mass
range which is most strongly affected by dynamical erosion.  The GCLF is 
to first order independent of metallicity and age.  Thus, globular
cluster formation seems to require only
a large input supply of gas clumped together into the SGMC-sized clouds 
inferred (see above) for the Milky Way.  
If this approach is basically correct, then
it is highly encouraging of the view that the sites of massive star cluster
formation happening now in (e.g.) NGC 1275 (Holtzman et al. 1992) or the
Antennae (Whitmore \& Schweizer 1995) do resemble what must have been
happening wholesale at much earlier epochs.

\noindent (2) The {\it metallicity distribution function}, which
repeatedly shows a bimodal form in giant ellipticals and spirals alike.
The metal-richer subpopulation is, invariably,
the more centrally concentrated.  The interpretation that these are the
signatures of two major, distinct epochs of cluster formation is now compelling.
But what drove these bursts, and why just two?  Answers are still not clear,
and the choice of mechanisms is now apparently larger than we had 
thought a few years ago.  These mechanisms can be thought of roughly in two
categories:  ``external'' 
(environmental effects from outside the galaxy in question), or ``internal''
(processes strictly within the galaxy and largely immune to environment).  

The most well known 
external mechanism is {\it mergers}, invoked as an important source of
globular cluster formation by Schweizer (1987) and Ashman \& Zepf (1992) and 
frequently mentioned since.  We now 
have clear-cut evidence that massive star clusters can and do form 
out of the shocked gas during mergers
(e.g. Whitmore \& Schweizer 1995; Schweizer et al.~1996; Holtzman 
et al.~1992, 1996).  Lee et al. (1997) and
Geisler et al.~(1996) favor the interpretation that the metal-richer cluster 
population in typical gE's was formed in a merger-stimulated burst. 
This approach may explain E galaxies with low specific
frequencies; but the gE's and cD galaxies with very high specific frequencies
still appear to need a qualitatively different approach.
Unless the cluster formation efficiency is phenomenally high 
during the merger, the colliding pre-cD `galaxies' must be so gas-rich to yield 
the requisite number of globular clusters that they must essentially 
be protogalactic clouds (Harris 1995).  Other, more detailed, difficulties
are discussed by Forbes et al. (1997a) and Geisler et al. (1996).
\begin{figure}
\plotfiddle{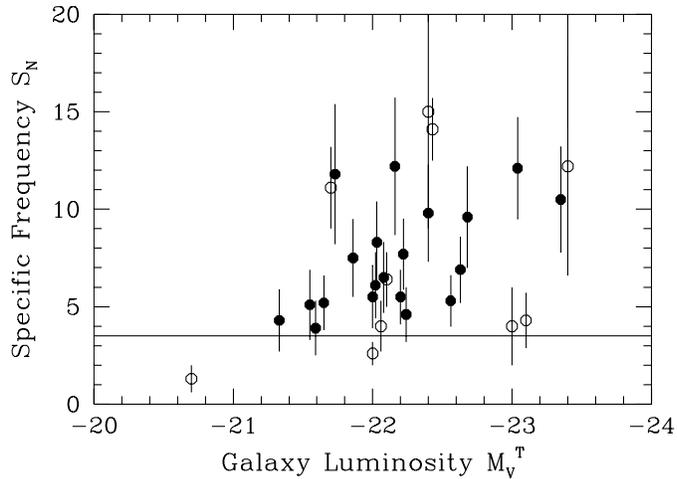}{2.50in}{0}{60}{60}{-200}{-130}
\caption{Specific frequencies $S_N$ (number of globular clusters per
unit galaxy luminosity) for cD-type galaxies (brightest cluster members).
Solid dots are from Blakeslee et al. (1997), and open circles
are other cD's from Harris et al. (1997b). More luminous cD's are 
preferentially in richer clusters, and have higher 
specific frequencies.  The line at $S_N = 3.5$ represents the average
for normal (non-cD) ellipticals.} \label{fig-4}
\end{figure}

The cD/BCG galaxies make up a subclass of special interest.
The new data of Blakeslee et al. (1997) add considerably to what we 
know of these centrally dominant systems (Figure~\ref{fig-4}).  
It is now clear that
$S_N \sim N_{cl}/L_{gal}$ varies closely with the depth of the surrounding
potential well (as measured by the velocity dispersion of the surrounding
galaxies, or the X-ray gas temperature). The bigger cD's
sit in deeper potentials and have more clusters per unit light.
West et al. (1995) propose that such galaxies sit amidst an extended halo
of {\it intragalactic clusters} belonging to the potential well of the surrounding
galaxy cluster. Grillmair et al.~(1994) and 
Kissler-Patig et al. (this conference) present tentative
evidence that such a population has been found around NGC 1399.
Furthermore, C\^ot\'e et al. (1997) show that the shape of the GCS metallicity
distribution in giant ellipticals can be matched if it is assumed that the
metal{\it-poor} clusters were acquired from tidally stripped smaller neighbors
(galaxy {\it harassment})
(though the high specific frequency and total cluster population present much
larger problems in this scenario).

Of the ``internal'' mechanisms, Forbes et al. (1997a) favor the 
more traditional (in some sense) view that the globular clusters
in normal E galaxies formed {\it in situ} in two or more major epochs 
of {\it local gaseous collapse and star formation}, regardless of 
outside influences.  They suggest that the high$-S_N$ cD-type ellipticals may then
have acquired large extra cluster populations by tidal stripping from
other neighboring galaxies.
The traces of these events are left in the GCS metallicity distribution and
space distribution.  
Later on, {\it dynamical erosion} -- tidal shocking, dynamical friction, stellar
evaporation -- will gradually remove clusters and reduce their masses; it
is intriguing to note that these effects may act in such a way as to preserve
the basic shape of the GCLF
that is built in at birth (McLaughlin \& Pudritz 1996; Vesperini 1997; 
Elmegreen \& Efremov 1997).
Finally, Harris et al.~(1997b) have noted the potential importance of 
{\it galactic winds}, particularly for the massive ellipticals in which
the first star formation burst is expected to be rapid.  That is, if indeed the
halo globular clusters form during the first Gyr, then they are likely to
have been affected strongly by the first SNII-driven galactic wind, which must
have been going on at very much the same time.  If the wind is capable of driving
out a large amount of the initial gas, then it is conceivable that the very high
specific frequencies in the cD galaxies are the visible result:  that
is, many clusters formed in the first burst, 
but before the remaining gas could 
continue well into its subsequent normal star formation, it was 
expelled, eventually leaving behind a large galaxy with an 
artificially high ratio of clusters to field stars.

For each of the individual mechanisms mentioned above, we can
point to a galaxy in which that particular process seems to be
the dominant one in the evolution.  But it appears that
no single approach is the answer to every situation,
and of the impressive list of processes capable of globally
influencing the characteristics of GCSs,  we are going to have to understand
them all.

\end{document}